\documentclass[aps,prl,10pt,twocolumn,superscriptaddress,longbibliography]{revtex4}

\usepackage{graphicx}
\usepackage{amssymb, amsmath,amsfonts}
\usepackage{color}
\usepackage{ulem}
\usepackage{bbm}
\usepackage{subfigure}
\usepackage{dcolumn}
\usepackage{mathtools}
\usepackage{pifont}
\usepackage{MnSymbol}

\def\I{\uppercase\expandafter{\romannumeral 1}}
\def\II{\uppercase\expandafter{\romannumeral 2}}
\def\III{{\uppercase\expandafter{\romannumeral 3}}}
\def\IV{{\uppercase\expandafter{\romannumeral 4}}}
\def\V{{\uppercase\expandafter{\romannumeral 5}}}
\def\VI{{\uppercase\expandafter{\romannumeral 6}}}
\def\VII{{\uppercase\expandafter{\romannumeral 7}}}
\def\i{\lowercase\expandafter{\romannumeral 1}}
\def\ii{\lowercase\expandafter{\romannumeral 2}}
\def\iii{{\lowercase\expandafter{\romannumeral 3}}}
\def\iv{{\lowercase\expandafter{\romannumeral 4}}}
\def\v{{\lowercase\expandafter{\romannumeral 5}}}
\def\vi{{\lowercase\expandafter{\romannumeral 6}}}
\def\vii{{\lowercase\expandafter{\romannumeral 7}}}
\def\nn{\nonumber\\}
\def\r135{RbV$_3$Sb$_5$}

\begin{document}

\title{First principles theory for the magnetic and charge instabilities  in AV$_3$Sb$_5$ systems}

\author{Hai-Yang Ma}
\affiliation{School of Physical Science and Technology, ShanghaiTech University, Shanghai 200031, China}
\affiliation{ShanghaiTech laboratory for topological physics, ShanghaiTech University, Shanghai 200031, China}



\author{Jianpeng Liu}
\email[]{liujp@shanghaitech.edu.cn}
\affiliation{School of Physical Science and Technology, ShanghaiTech University, Shanghai 200031, China}
\affiliation{ShanghaiTech laboratory for topological physics, ShanghaiTech University, Shanghai 200031, China}

\begin{abstract}
Vanadium-based materials AV$_3$Sb$_5$ (A=K, Rb, Cs) with layered kagome lattice structures have drawn great attention recently due to the discoveries of topologically nontrivial band structures, charge density wave states, giant anomalous Hall effect, as well as unusual superconducting phase at low temperatures. In this work, we theoretically study the magnetic and charge instabilities for this class of materials based on first principles calculations. We develop a method to calculate the generalized susceptibility tensor defined in the sublattice-orbital-spin space, with the effects of Coulomb interactions treated by generalized random phase approximation (RPA). The RPA susceptibility calculations indicate that there are three leading ferromagnetic instability modes at $\Gamma$ point, which are further verified by unrestricted self-consistent Hartree-Fock calculations including both the on-site and inter-site Coulomb interactions. The inclusion of inter-site  interactions tend to suppress the spin ferromagnetism due to charge transfer from the V to Sb sites, leading to weak spin magnetic moments $\sim 0.1\mu_{\textrm{B}}$ per V atom with small intrinsic anomalous Hall conductivity. Current loops can be generated in such weak spin ferromagnetic states as a result of spin-orbit coupling effects, giving rise to orbital magnetic fluxes.  The electronic structures in the ferromagnetic states are significantly reconstructed which have nearly compensated electron and hole carriers from two bands.   On the other hand, we do not find any diverging instability mode at  $M$ point driven by electron-electron Coulomb interactions. First principles phonon calculations indicate that there are  unstable phonon modes which tend to drive the system into an inverse star-of-David structure. Our results indicate that there may be separate phase transitions  in the magnetic and charge channels in the system. 

\end{abstract}


\maketitle
Quantum materials with layered kagome structures are of great interest due to the frustrated lattice geometry, which can give rise to interesting non-interacting band structures with both flat bands and Dirac cones \cite{balents-prb08}. The nontrivial Berry phase associated with the Dirac fermions and the vanishing kinetic energy of the flat bands make the kagome lattice system an ideal platform to study the interplay between  band topology and strong electron-electron ($e$-$e$) correlations, which can lead to various exotic quantum states \cite{wen-prl11,thomale-kagome-prl13,wang-kagome-prb13,thomale-kagome-prb12,li-kagome-prb12,yin2018giant,ye2018massive,liu2018giant,yin2020quantum,ko2009doped,balents2010spin,kiesel2013unconventional}.  Recently, vanadium-based layered kagome materials AV$_3$Sb$_5$ (A=K, Rb, Cs) have been successfully synthesized \cite{ortiz2019new,yang2020giant}, which exhibit various intriguing properties including  charge density wave states \cite{jiang2021unconventional,li2021no,Yu2021Concurrence,liang2021three,zhao2021cascade,li2021absence,wang2021electronic,ortiz2021fermi,nakayama2021multiple,li2021rotation,kang2021twofold,xiang2021nematic,wang2021distinctive,liu2021temperature,cho2021emergence,Zhou2021Origin,xie2021electron,wulferding2021fermi,wang-optics-prb21,li-thickness-arxiv21},  the signature of time-reversal symmetry breaking \cite{yang2020giant,mielke2021time,yu2021evidence,wu2021large}, as well as unusual superconductivity \cite{ortiz2021superconductivity,yin2021superconductivity,ortiz2020cs,chen2021double,zhang2021pressure,chen2021highly,Li2021Observation,zhao2021nodal,gupta2021microscopic,Xu2021Multiband,yu2021unusual,lou2021charge,ni2021anisotropic,li-thickness-arxiv21,chen2021roton,yin2021strain}. 

These experimental observations have boosted intensive theoretical interest \cite{balents-instability-prb21,wu2021nature,denner2021analysis,miao2021geometry,lin2021complex,feng2021chiral,consiglio2021van,ye2021structural,Tan2021Charge,gu2021gapless,wang2021origin,jiang-kagome-review,feng-prb21}. First principles calculations reveal that the band structures of AV$_3$Sb$_5$ are topologically nontrivial with nonzero $\mathbbm{Z}_2$ indices and Dirac-like linear dispersions, and there are multiple van Hove singularties near the Fermi level \cite{ortiz2020cs}. Theoretical studies  on the interaction-driven Fermi-surface instabilities based on simplified model Hamiltonians predict the presence of ``chiral flux modes" and various other instability channels, which provide a unified picture for the formation of charge density wave and time-reversal breaking states \cite{wu2021nature,feng2021chiral,gu2021gapless}.

However, recent phonon calculations indicate the presence of unstable phonon modes in the system, which provides another possible scenario for the emergence of CDW phase \cite{Tan2021Charge,wang2021origin,cho2021emergence,ye2021structural}. There are also a lot of sublattice, orbital, and spin degrees of freedom in each primitive cell, and $e\rm{-}e$ interactions can lead to incredibly complicated symmetry-breaking states in the charge-orbit-spin space. In order to unambiguously identify the leading  Fermi-surface instability modes and shed light on the nature of the charge and magnetic properties in this system, we calculate the generalized two-particle susceptibility tensor defined the sublattice-orbital-spin space based on realistic 60-band Wannier tight-binding models generated from first principles calculations, including the effects of $e\textrm{-}e$ Coulomb interactions as treated by random phase approximation \cite{ren2012random,korshunov2014superconducting}.  Taking the leading eigenmodes of the generalized susceptibility tensor as an initial ansatz, we further perform un-restricted self consistent Hartree Fock calculations for both primitive cell and $2\times 2 \times 1$ inverse-star-of-David (ISD) structures of AV$_3$Sb$_5$ (A=K, Rb, Cs) system,  including both on-site Kanamori-type interactions  and inter-site density-density interactions, with nonlocal exchange effects being treated exactly. Through these calculations, we find that the ground-state manifold of the system consists of three nearly degenerate spin ferromagnetic states. The inclusion of inter-site Coulomb interactions tend to suppress the spin ferromagnetism, leading to weak spin magnetic moments $\sim\!0.1\,\mu_{\textrm{B}}$ per V atom. Charge current loops (with current amplitudes $\sim20\textrm{-}100\,$nA) can be generated in such weak spin ferromagnetic states as a result of spin-orbit coupling effects, giving rise to orbital magnetic fluxes threaded through the kagome plane.  We further study the phonon properties of the AV$_3$Sb$_5$ (A=K, Rb, Cs) system, and find unstable phonon modes that can give rise to the ISD crystal structure, which implies a phonon-driven mechanism for the CDW phase. Our results indicate that there are separate magnetic and charge instabilities in the system, and that there may be additional magnetic phase transition within the ISD CDW phase.

\begin{figure}[!htbp]
\includegraphics[width=3.5in]{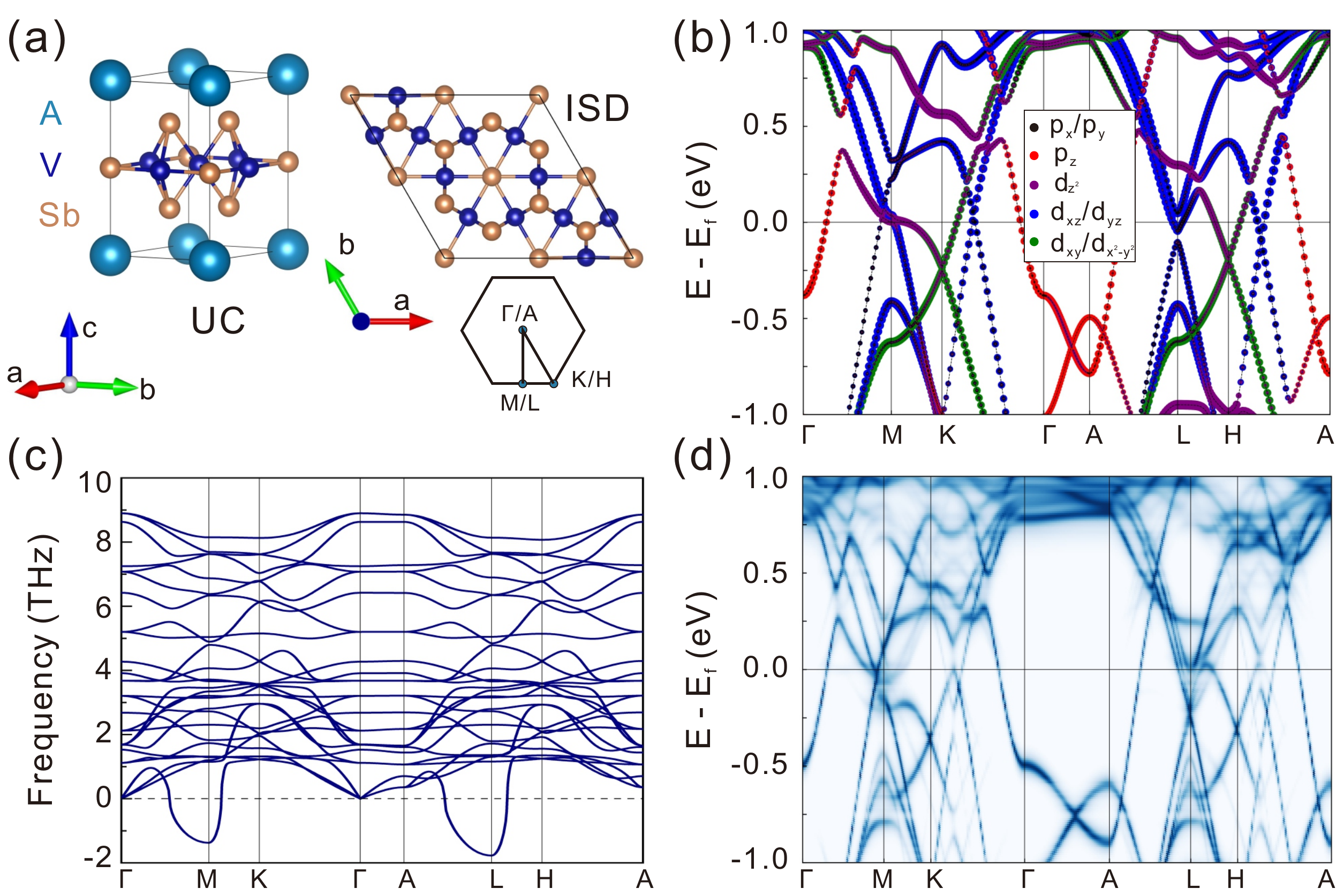}
\caption{~\label{fig1} (a) Lattice structures of the normal phase (pristine structure) and the CDW phase (inverse-star-of-David structure) of AV$_{3}$Sb$_{5}$, A = K, Rb or Cs. (b) Band structures for the normal phase of RbV$_{3}$Sb$_{5}$ projected onto V $d$ and Sb $p$ orbitals. (c) Phonon spectrum for the normal phase of RbV$_{3}$Sb$_{5}$. Two unstable modes can be found at the $M$ and $L$ points, indicating structure instability. (d) Unfolded spectral functions of the CDW phase of RbV$_{3}$Sb$_{5}$. }
\end{figure}

\paragraph{Preliminaries \textemdash}

The AV$_3$Sb$_5$ (A=K, Rb, Cs) system crystallizes in a layered kagome lattice structure as shown in Fig.~\ref{fig1}(a). In each primitive cell there are three V sublattices forming kagome plane, within which one Sb atom is intercalated into each primitive cell. There are two layers of Sb atoms forming honeycomb lattices above and below the vanadium kagome planes, forming V-Sb hexahedra frameworks. The alkali metal are further intercalated between the V-Sb layers, donating electrons and making this system highly metallic. In Fig.~\ref{fig1}(b) we show the band structure of RbV$_3$Sb$_5$ projected onto V $3d$ orbitals and Sb $5p$ orbitals in the pristine lattice structure which are calculated by first principles density functional theory (DFT). Clearly there are Dirac-like dispersions near $K$ and $H$ points contributed by V $d$ orbitals,  van Hove singularities near $M$ point contributed by both V $d$ and Sb $p$ orbitals, as well as electron pockets centered at $\Gamma$ which are mostly contributed by Sb $p$ orbitals. Such complicated Fermi surfaces and orbital characters make it difficult to realistically describe the electronic structure of this system using a simplified lattice model with few orbitals. 
 In Fig.~\ref{fig1}(c) we present the phonon dispersions calculated by DFT, where we see two unstable phonon modes at the $M$ and $L$ points, consistent with previous reports. A linear combination of the unstable modes at the three equivalent $M$ points result in an inverse-star-of-David structure with $2\times 2\times 1$ supercell as shown in the right panel of Fig.~\ref{fig1}(a). The band structures of the system are strongly re-constructed in the ISD lattice structure as clearly shown by the spectral function unfolded to the original primitive Brillouin zone (Fig.~\ref{fig1}(d)), and new van Hove singularities emerge as a result of the lattice deformation \cite{cho2021emergence}.

\paragraph{Fermi surface instabilities \textemdash}

The presence of van Hove singularities both in the pristine and in the ISD structures indicate that the non-interacting Fermi surfaces may be unstable to $e\textrm{-}e$ Coulomb interactions. In order to study the interaction effects on the complicated Fermi surfaces with diverse orbital characters as shown in Fig.~\ref{fig1}(b), we  construct a realistic tight-binding model in the Wannier-function basis using the Wannier90 code package \cite{mostofi2008wannier90}. To be specific, the Bloch functions generated from DFT calculations are projected onto 60 spinor Wannier functions (for each pristine primitive cell) including all the V 3$d$ orbitals and Sb $5p$ orbitals, from which we obtain the following Wannier tight-binding model denoted by $H_0$:
\begin{equation}
H_0=\sum_{i\alpha\sigma,j\beta\sigma'} h_{i\alpha\sigma,j\beta\sigma'}\,\hat{c}^{\dagger}_{i\alpha\sigma}\, \hat{c}_{j\beta\sigma'}
\label{eq:tb}
\end{equation}
where the sets of indices $\{i,j\}$,
$\{\alpha,\beta\}$, $\{\sigma,\sigma'\}$ denote, in turn, 
all the V and Sb atomic sites, orbitals and spin degrees of freedom, and $h_{i\alpha\sigma,j\beta\sigma'}$ denotes the real-space hopping amplitude including spin-orbit coupling effects . $\hat{c}^{\dagger}_{i\alpha\sigma}$ and $\hat{c}_{j\beta\sigma'}$ are the electron creation and annihilation operators.
The Wannier interpolated band structures can exactly reproduce those obtained from DFT calculations within an energy window extending from 2\,eV below to 2\,eV above the Fermi energy.  We include the on-site multi-orbital Coulomb interactions of the Kanamori type for the V 3$d$ orbitals    \cite{kanamori}, 
\begin{align}
H_{\textrm{K}}=&U\sum_{i,\alpha}\hat{n}_{i\alpha\uparrow}\hat{n}_{i\alpha\downarrow}+
U'\sum_{i,\alpha <\beta,\sigma,\sigma'}\hat{n}_{i\alpha\sigma}\hat{n}_{i\beta\sigma'}\;\nn
&-J_{\textrm{H}}\sum_{i,\alpha < \beta,\sigma,\sigma'}\hat{c}^{\dagger}_{i\alpha\sigma}
\hat{c}^{\vphantom\dagger}_{i\alpha\sigma'}\hat{c}^{\dagger}_{i\beta\sigma'}\hat{c}^{\vphantom\dagger}_{i\beta\sigma}\;\nn
&+J_{\textrm{P}}\sum_{i,\alpha < \beta,\sigma}
\hat{c}^{\dagger}_{i\alpha\sigma}\hat{c}^{\dagger}_{i\alpha-\sigma}\hat{c}^{\vphantom\dagger}_{i\beta\sigma}\hat{c}^{\vphantom\dagger}_{i\beta -\sigma}\;,
\label{eq:kanamori}
\end{align}
where $U$ and $U'$ are the intra-orbital and inter-orbital direct Coulomb interactions. 
$J_{\textrm{H}}$ and $J_{\textrm{P}}$ denote the Hunds' coupling 
and pair hoppings respectively, and $\hat{n}_{i\alpha\sigma}=\hat{c}^{\dagger}_{i\alpha\sigma}\hat{c}_{i\alpha\sigma}$ is the  density operator. We set $U\!=\!3.6\,$eV and $J\!=\!J_{\textrm{H}}\!=\!0.72\,$eV throughout this paper, and we let $U'\!=\!U-J$ and $\!J_{\textrm{P}}\!= 0$ such that Eq.~(\ref{eq:kanamori}) has full rotational invariance \cite{georges2013strong}.
We further consider the density-density inter-site interactions between the neighboring V-Sb sites and V-V sites,  the interaction amplitudes are denoted by $V$ and $W$ respectively,
\begin{align}
&H_{\textrm{V-Sb}}=V\,\sum_{\langle i  j \rangle}\sum_{\alpha\beta\sigma\sigma'}\,\hat{c}^{\dagger}_{i\alpha\sigma}\hat{c}^{\dagger}_{j\beta\sigma'}\hat{c}_{j\beta\sigma'}\hat{c}_{i\alpha\sigma}\;\nn
&H_{\textrm{V-V}}=W\,\sum_{\llangle ij \rrangle}\sum_{\alpha\beta\sigma\sigma'} \,\hat{c}^{\dagger}_{i\alpha\sigma}\hat{c}^{\dagger}_{j\beta\sigma'}\hat{c}_{j\beta\sigma'}\hat{c}_{i\alpha\sigma}\;,
\label{eq:intersite}
\end{align}
where ``$\sum_{\langle ij \rangle}$" means a summation over all the first-neighbor V-Sb sites (including both intralayer ones and interlayer ones), and the symbol ``$\sum_{\llangle ij \rrangle}$" means summing over all the in-plane nearest neighbor V-V sites. In this work $V$ and $W$ are treated as two free parameters which vary from 0 to 1\,eV. The full Hamiltonian is then $H_0+H_{\textrm{K}}+H_{\textrm{V-Sb}}+H_{\textrm{V-V}}$. 

As discussed above, the system has a quite complicated primitive cell with multiple sublattice, orbital, and spin degrees of freedom. For the 60-band tight-binding model, there are 3600 independent matrix elements of the density operator $\rho_{i\alpha\sigma,j\beta\sigma'}=\langle \hat{c}^{\dagger}_{i\alpha\sigma} \hat{c}_{j\beta\sigma'} \rangle$ (consider $i,j$ as sublattice sites within the same primitive cell),  each of which can be considered as an ``order parameter" that break certain symmetries of the system and bring about Fermi surface instability. The actual ground state may be characterized by a complicated linear combination of the density matrix elements, which is difficult to guess a priori due to such complexity. Therefore, in order to  find the leading Fermi-surface instability mode, we propose to calculate the interaction-renormalized generalized susceptibility tensor defined in the sublattice-orbital-spin space, the tensor element of which is denoted as $\chi(\mathbf{q})_{\mu\nu,\mu'\nu'}$, with $\mu\equiv\{i\alpha\sigma\}$ being  a composite index referring to all the sublattice, orbital, and spin degrees of freedom (so do $\mu',\nu$, and $\nu'$). Suppose there is  a weak static field with wavevector $\mathbf{q}$  that is only coupled with some given element of the density operator 
$\rho_{\mu'\nu'}(\mathbf{q})$ (denoted as $A_{\mu'\nu'}(\mathbf{q})$) is applied to the system, a change of density matrix element would be induced by the field, i.e.,
$\delta\rho_{\mu\nu}(\mathbf{q})=\chi(\mathbf{q})_{\mu\nu,\mu'\nu'}\,A_{\mu'\nu'}(\mathbf{q})$. One can further consider a field $\mathbf{A}^{\kappa}$ that is only coupled to (or ``parallel" to) the $\kappa$th eigenmode $u_{\mu\nu}^{\kappa}(\mathbf{q})$ of the susceptibility tensor, i.e., $\mathbf{A}^{\kappa}(\mathbf{q})\sim u_{\mu\nu}^{\kappa}(\mathbf{q})$, then the system would respond to such a  field along the eigenvector direction via
\begin{equation}
\delta\rho_{\mu\nu}(\mathbf{q}) \sim \lambda^{\kappa}(\mathbf{q}) u_{\mu\nu}^{\kappa}(\mathbf{q})
\end{equation}
where $\lambda^{\kappa}(\mathbf{q})$ and $u_{\mu\nu}^{\kappa}(\mathbf{q})$ are the eigenvalues and eigenvectors of the susceptibility tensor $\chi(\mathbf{q})$. If any of the eigenmodes has a diverging eigenvalue $\lambda(\mathbf{q})^{\kappa}$, it means that the system would develop a spontaneous order parameter characterized by the diverging eigenmode $u_{\mu\nu}^{\kappa}(\mathbf{q})$ upon the application of an infinitesimal field, and would enter a spontaneous symmetry-breaking state through a second order phase transition. Note that similar idea has been used in the studies of surface-state instabilities of nodal-line semietals \cite{liu-prb17} and octupolar instabilities of RuCl$_3$ etc. \cite{xu-prl21,motome-prb15}

\begin{figure}[!htbp]
\includegraphics[width=3.5in]{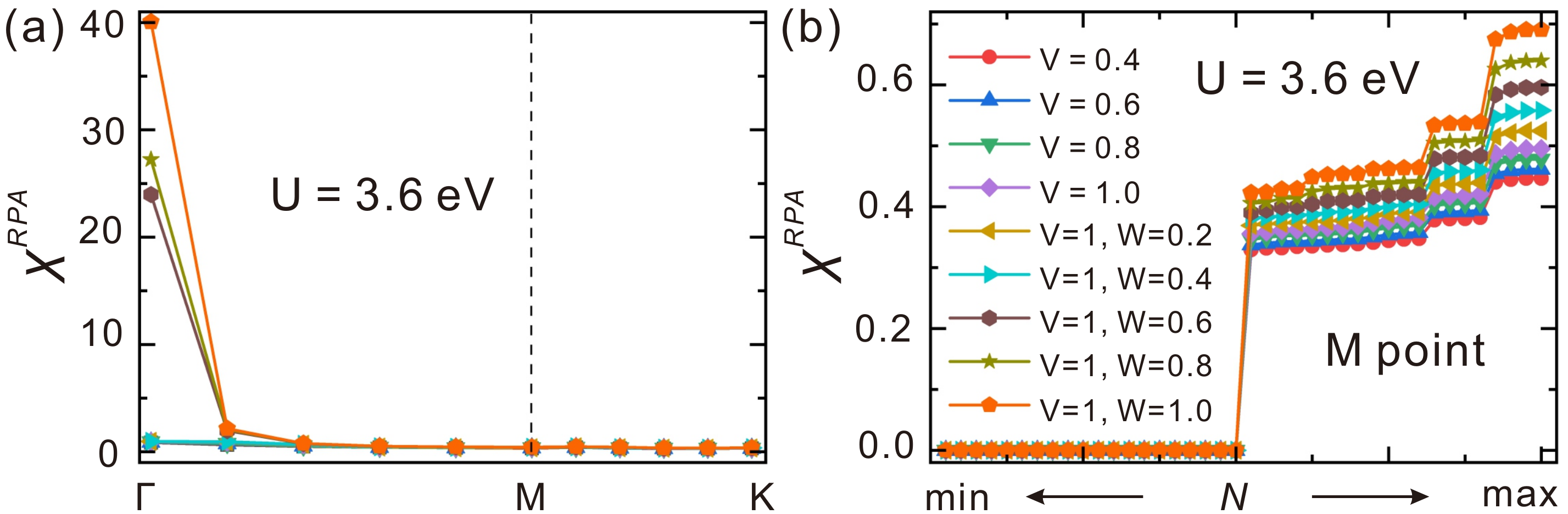}
\caption{~\label{fig2} (a) Eigenvalues of the generalized RPA susceptibility tensor of Rb$V_{3}$Sb$_{5}$ (with on-site interaction only) , plotted along a high-symmetry $\mathbf{q}$ path. Only the 10 leading eigenvalues are shown.  (b) The eigenvalues of the RPA susceptibility at $M$ point for different V-Sb ($V$) and V-V ($W$) inter-site interaction strengths. The on-site interaction parameters are taken to be $U\!=\!3.6\,$eV and $J\!=\!0.72\,$eV.}
\end{figure}

Following the above argument, we calculate the generalized susceptibility tensor for the AV$_3$Sb$_5$ (A=K, Rb, Cs) system based on the Wannierized tight-binding Hamiltonian given by Eqs.~(\ref{eq:tb})-(\ref{eq:intersite}), and the effects of the on-site and inter-site Coulomb interactions are treated by random phase approximation (RPA), the details of which are presented in Supplementary Information \cite{supp_info}. For clarity's sake, in main text we only present the results for RbV$_3$Sb$_5$, and the results of the other two compounds are given in Supplementary Information \cite{supp_info}. In Fig.~\ref{fig2}(a) we present the 10 eigenmodes with maximal eigenvalues (of the RPA susceptibility) as a function of the wavevector $\mathbf{q}$, with on-site interaction only and the inter-site interactions being temporarily turned off. We see that there are three diverging modes only at $\Gamma$ point. These diverging modes at $\Gamma$ point turn out to be three spin magnetic modes. The inclusion of inter-site interactions does not change the diverging nature of the three spin ferromagnetic modes.
On the other hand, since the system undergoes a CDW transition forming an ISD structure around 100\,K, it is worthwhile to study the microscopic origin of such CDW transition. In such a context, we also carefully study the eigenmodes of the RPA susceptibility tensor at the three $M$ points, and find that there are no notable unstable modes. We further include V-Sb and V-V inter-site interactions, and find that the eigenvalues of the susceptibility tensor at $M$ points only increase by small amounts due to the inclusion of intersite interactions, as clearly shown in Fig.~\ref{fig2}(b), where the different markers represent data sets for different choices of $V$ and $W$ values.  Therefore, the CDW phase is likely driven by phonon instability and electron-phonon couplings instead of $e$-$e$ interactions.

\begin{figure}[!htbp]
\includegraphics[width=3.5in]{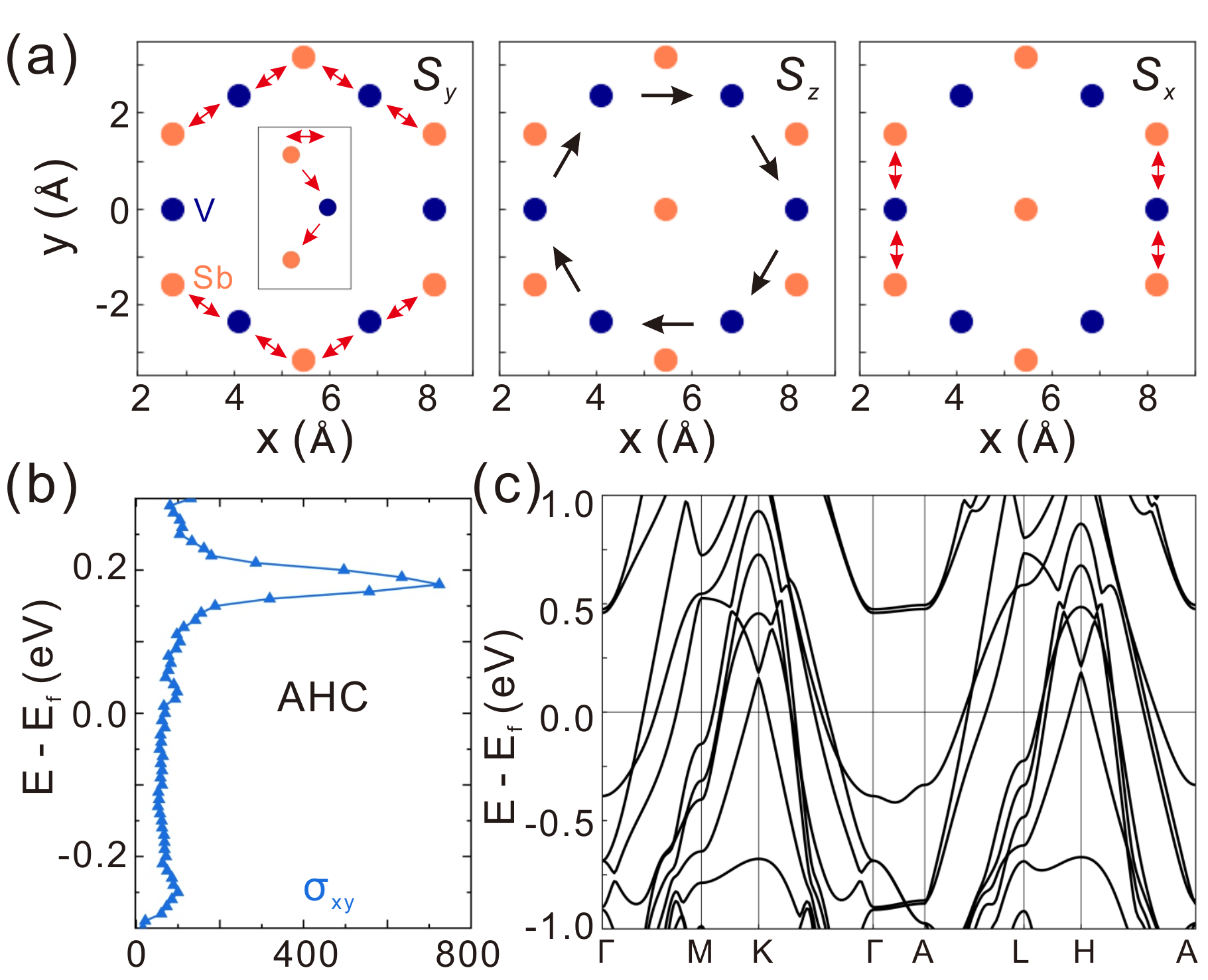}
\caption{~\label{fig3} (a) Schematic illustration of the inter-site currents for the three spin ferromagnetic states in the pristine lattice structure.  The black arrows denote the intra-plane currents between neighbouring V-V sites, while the double red arrows denote the inter-layer currents between the V atom and the neighbouring Sb atoms below and above the V kagome plane. (b) Calculated anomalous Hall conductivity of the $FM_z$ phase with magnetization primarily pointing to the crystalline $z$ direction. (c) Hartree-Fock band structures of the FM$_z$ phase in the pristine structure. Only on-site interactions are taken into account in the results presented in (a)-(c). }
\end{figure}

\paragraph{Hartree-Fock calculations \textemdash}
In order to verify the conclusions obtained from the susceptibility calculations, we further perform un-restricted self-consistent Hartree-Fock (HF) calculations including both the on-site Kanamori interaction (Eq.~(\ref{eq:kanamori})) and the inter-site interactions (Eq.~(\ref{eq:intersite})), with the nonlocal exchangee effects being treated exactly.  We first perform HF calculations for the pristine primitive cell, and we take the three leading instability modes at $\Gamma$ point (Fig.~\ref{fig2}(a)) as initial anstza for the density matrices in the HF calculations. The HF ground states turn out to be three spin ferromagnetic states with the magnetizations primarily pointing along the crystalline $x$, $y$, and $z$ directions with slight cantings, which are denoted as FM$_x$, FM$_y$, and FM$_z$ states respectively. With on-site interactions only ($U\!=\!3.6\,$eV, $J\!=\!0.72\,$eV, $W\!=\!V\!=\!0$), the spin magnetization is as large as $\sim\!0.6  \,\mu_{\textrm{B}}$ per V atom. Real-space current loops are generated in the ferromagnetic states as a result of spin-orbit coupling, which are shown in the three panels of Fig.~\ref{fig3}(a) for  the FM$_y$, FM$_z$, and FM$_x$ phases. The red (double) arrows indicate interlayer V-Sb currents, while the black arrows indicate V-V in-plane currents. The maximal current amplitudes $\sim 800\,$nA, generating an orbital moment $\sim\!0.2\mu_{\textrm{B}}$, e.g., for the hexagonal current loop in the middle panel of Fig.~\ref{fig3}(a). The filling dependence of the calculated anomalous Hall conductivity (AHC) $\sigma_{xy}$ in the FM$_z$ state is shown in Fig.~\ref{fig3}(b), where we see a peak of $\sigma_{xy}\!\sim\!800\,$S/cm when the Fermi energy is assumed to be $\sim\!0.15\,$eV above the physical Fermi energy; otherwise $\sigma_{xy}$ is quite small $\sim\!10\textrm{-}50\,$S/cm. The HF band structures (with on-site interactions only) are shown in Fig.~\ref{fig3}(c).

The results presented in Fig.~\ref{fig3} are generated for the case with on-site interactions only. The inclusion of inter-site interactions, e.g., with $V\!=\!1\,$eV and $W\!=\!0$, would significantly suppress the spin ferromagnetism, resulting in a spin moment $\sim\!0.06\,\mu_{\textrm{B}}$ per V atom, and the corresponding AHC is 
two orders of magnitudes smaller $\sim\! 0.5-5\,$S/cm. 
This can be interpreted from the charge transfer induced by the inter-site Sb-V Coulomb interactions. Once the inter-site interaction is taken into account, charges tend to transfer from the V to the Sb sites to minimize the Hartree energy, resulting in a charge occupation $\sim\!0.3\,$ per V atom for $V\!=\!\,1$eV, which is reduced by a factor of 6 compared  with the case when $V\!=\!0$. As a result, the spin magnetization driven by the strong on-site Coulomb interactions of the V sites is also significantly diminished.
Moreover, the energies differences of the three FM states are incredibly close to each other $\sim0.01\,$meV, which imply strong magnetic fluctuations in real space. Therefore, the giant anomalous Hall effect observed in experiments \cite{yang2020giant} likely arises from skew scatterings due to strong real-space magnetic fluctuations \cite{nagaosa-skew-prb21}. 
 The detailed results for the HF calculations including inter-site interactions with pristine primitive cell are presented in Supplementary Information \cite{supp_info}.

%

\begin{figure}[!htbp]
\includegraphics[width=3.5in]{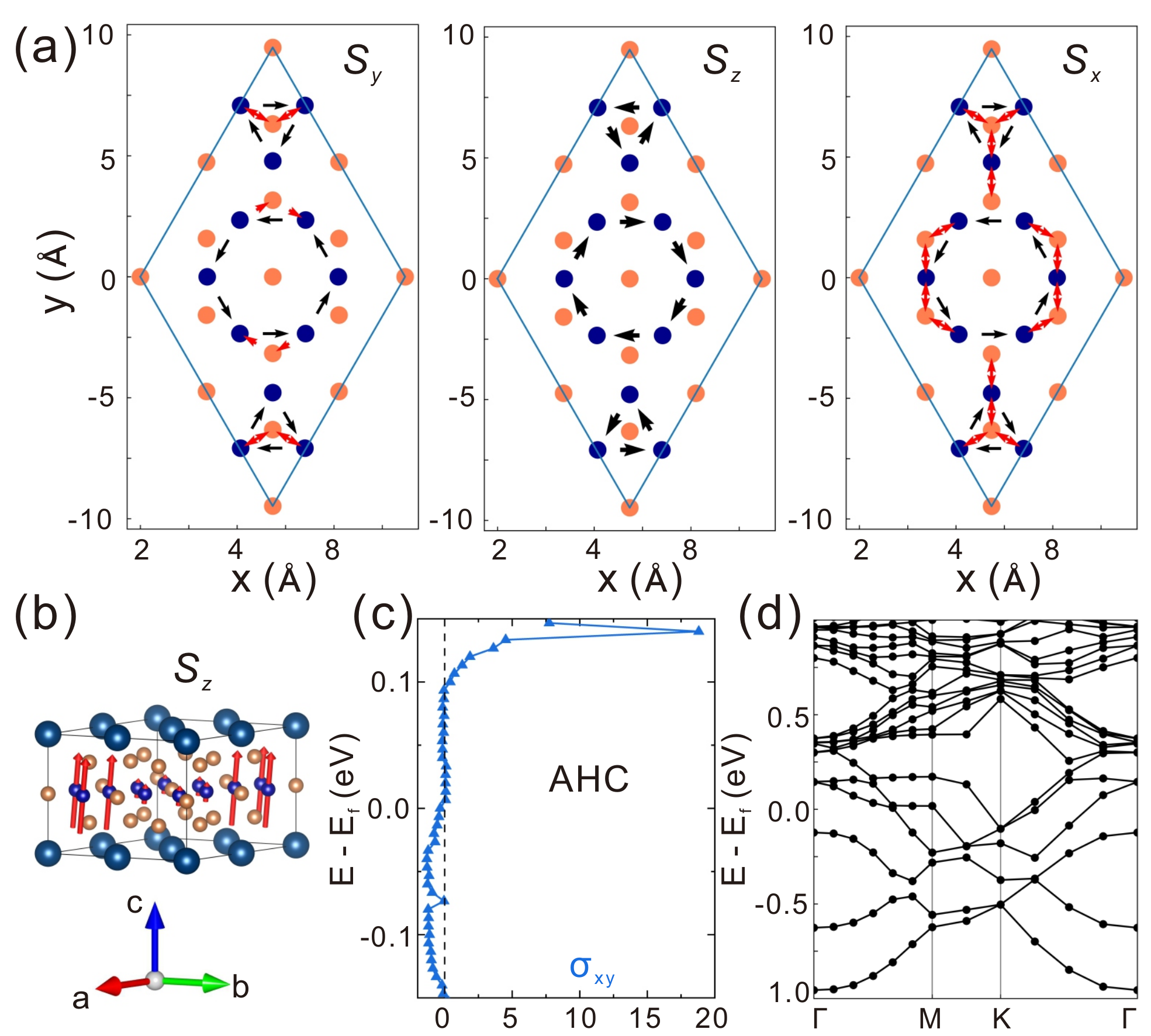}
\caption{~\label{fig4} (a) Schematic illustration of the inter-site currents for the three spin ferromagnetic states of RbV$_3$Sb$_5$ in the $2\times 2\times 1$ ISD structure, including both on-site interaction and V-Sb inter-site interaction ($V$), with $V=1.0$\,eV.(c) Magnetic structure of the FM$_z$ phase in the ISD lattice structure, with the arrows denoting the amplitudes and directions of the spin magnetization at different sites. (c) Calculated anomalous Hall conductivity of the FM$_z$ phase in the ISD structure. (d) Hartree-Fock band structures of the FM$_z$ phase in the ISD structure. }
\end{figure}

We have also performed un-restricted self-consistent HF calculations for the $2\!\times\!2\!\times\!1$ ISD lattice structure with a realistic tight-binding model including 240 Wannier orbitals, including both the on-site and inter-site interactions as given by Eqs.~(\ref{eq:tb})-(\ref{eq:intersite}), and the non-local exchange effects are treated exactly. We take the three spin ferromagnetic modes at $\Gamma$ point and four leading modes at $M$ point (obtained from RPA susceptibility calculations) as our initial ansatz for the HF calculations. The ground states, again, involve three nearly degenerate spin ferromagnetic states with their magnetizations pointing along the crystalline $x$, $y$, and $z$ directions respectively, with the energy difference $\sim 0.01\,$meV. Similar to the case of pristine primitive cell, the magnetizations in the $2\!\times\!2\!\times\!1$ ISD structure are on the order of $1\,\mu_{\textrm{B}}$ if only on-site interactions are included, and can be suppressed by one order of magnitude due to the inclusion of inter-site interactions. Interestingly, in the $2\!\times\!2\!\times\!1$ ISD structure, the magnetization distribution is inhomogeneous within the ISD primitive cell: the spin magnetizations are mostly concentrated at the six V atoms forming two contracted kagome triangles, and the other six V atoms forming a hexagon have nearly vanishing spin magnetizations, as schematically shown in Fig.~\ref{fig4}(b) for the FM$_z$ phase. 

As in the case of the pristine cell, current loops are generated in the FM states as a result of spin-orbit coupling. The current-loop patterns in the three FM states are shown in Fig.~\ref{fig4}(a) for the FM$_y$, FM$_z$, and FM$_x$ states respectively. For example, for the FM$_z$ state, there are three current loops: two of them are counter-clockwise circulating around the two kagome triangles, and the other one is clockwise circulating around the hexagon. With the intersite interaction $V\!=\!1\,$eV, the current amplitudes are very small $\sim 80\,$nA, which can generate a small orbital moment $\sim 0.02\,\mu_{\textrm{B}}$ for each current loop. We have further calculated the anomalous Hall conductivities of the three FM states in the ISD structure. The inclusion of V-Sb inter-site Coulomb interactions significantly suppress ferromagnetism and anomalous Hall effect (AHE), and the calculated AHC $\sigma_{xy}$ in the FM$_z$ state $\sim 1-20\,$S/cm as shown in Fig.~\ref{fig4}(c). Again, this indicates an incredibly small intrinsic contribution to AHE, and the giant AHC measured in experiments \cite{yang2020giant} are likely due to skew scatterings from the strong magnetic fluctuations in real-space  by virtue of the near degeneracy of the three ferromagnetic states \cite{nagaosa-skew-prb21}. In Fig.~\ref{fig4}(d), we have shown the calculated HF band structures of the $2\!\times\!2\!\times\!1$ ISD structure (plotted in the supercell Brillouin zone) with intersite interaction V=1.0\,eV, from which we see a strong renormalization effects on the electronic structures due to the inclusion of nonlocal interactions. Notably, the Fermi surfaces are mostly contributed by two bands with nearly compensated electron-type and hole-type carriers due to the onset of magnetism, which is consistent with the Hall-effect measurements below $\sim 30\,$K \cite{yang2020giant}. 
Our results indicate that there may be two transitions in the AV$_3$Sb$_5$ (A=K, Rb, Cs) system, the first one is the phonon-driven CDW transition, and the other is the magnetic transition (likely with lower $T_{c}$) marked by the emergence of giant anomalous Hall effect  and the reconstruction of Fermi surfaces due to the onset of weak spin ferromagnetism. 

\paragraph{Summary \textemdash}
To summarize, in this work we have comprehensively studied the magnetic and charge instabilities of the AV$_3$Sb$_5$ system based on first principles tight-binding models including effects of both on-site and inter-site Coulomb interactions. Results from both RPA susceptibility calculations and un-restricted Hartree-Fock calculations indicate that the leading Fermi-surface instabilities driven by $e$-$e$ interactions are three nearly degenerate spin ferromagnetic modes regardless whether the system is in the pristine structure or ISD structure.
The competition between the nearly degenerate ferromagnetic states can lead to strong real-space magnetic fluctuations, which may be responsible for the  giant anomalous Hall effect observed in experiments. Current loops can be generated in the weak spin ferromagnetic states as a result of spin-orbit coupling effects, giving rise to spin-dependent orbital magnetic fluxes.  The electronic structures in the ferromagnetic states are significantly reconstructed which have nearly compensated electron and hole carriers from two bands, which may explain the recent transport data at low temperatures.   On the other hand, we do not find any diverging instability mode at  $M$ point driven by electron-electron Coulomb interactions. The unstable phonon modes at $M$ and $L$ points indicate that the CDW state in the system is likely driven by phonon instability and electron-phonon couplings. Our results imply separate magnetic and charge instabilities in the AV$_3$Sb$_5$ (A=K, Rb, Cs) system, and the system may undergo an additional magnetic phase transition marked by the onset of anomalous Hall effect and reconstruction of Fermi surfaces at a critical temperature lower than the CDW transition temperature. Our work sheds light on the puzzling magnetic and charge properties of the V-based kagome metal systems, and will provide useful guidelines for future studies.
\acknowledgements
\paragraph{Acknowledgements. \textemdash} We thank Yueshen Wu, Jun Li,  Kun Jiang, and Zheng Han for valuable discussions. This work is supported by  the National Science Foundation of China (grant no. 12174257), the National Key R \& D program of China (grant no. 2020YFA0309601), and the start-up grant of ShanghaiTech University. 

\bibliography{AV3Sb5}
\end{document}